\documentclass[11pt]{article}
\pdfoutput=1

\usepackage{jcappub}
\usepackage[table]{xcolor}
\usepackage{makecell}
\usepackage{caption}
\usepackage{subcaption}
\usepackage{hyperref}
\usepackage{graphicx}
\usepackage{amsmath}
\usepackage{mathtools}
\usepackage{physics}
\usepackage{siunitx}
\usepackage{tablefootnote}
\usepackage{soul}
\usepackage{color}
\usepackage{verbatim}
\usepackage{float}
\usepackage{bm}
\usepackage{ulem}
\usepackage{orcidlink}
\usepackage[nameinlink, capitalise]{cleveref}
\usepackage{braket}
\usepackage{dirtytalk}
\usepackage{mathtools}
\usepackage{amsmath}

\usepackage{tikz} 
\usepackage{tikz-feynman}
\tikzfeynmanset{compat=1.1.0}
\usepackage{feynmp}
\usetikzlibrary{shapes.geometric, backgrounds, arrows, calc, fit}
\tikzstyle{process} = [rectangle, rounded corners, minimum width=8cm, minimum height=1cm, text centered, draw=black, fill=blue!30]
\tikzstyle{arrow} = [thick,->,>=stealth]
\usepackage{underscore}

\usepackage{color,appendix,latexsym,amsmath,amssymb,graphicx,booktabs,epsfig,hyperref,url}

\numberwithin{equation}{section}
\usepackage[english]{babel}

\usepackage{letltxmacro}
\LetLtxMacro{\originaleqref}{\eqref}
\usepackage[sorting=none]{biblatex} 
\addto\extrasenglish{%
}

\definecolor{MyBlue}{rgb}{0.15,0.15,0.70}
\hypersetup{
colorlinks=true,
citecolor=MyBlue,
linkcolor=MyBlue,
urlcolor=MyBlue
}

\definecolor{orange}{rgb}{0.98, 0.6, 0.01}
\definecolor{darkolivegreen}{rgb}{0.33, 0.42, 0.18}
\definecolor{tealblue}{rgb}{0.21, 0.46, 0.53}

\usepackage{xspace}

\usepackage{listings}
\definecolor{codegreen}{rgb}{0,0.6,0}
\definecolor{codegray}{rgb}{0.5,0.5,0.5}
\definecolor{codepurple}{rgb}{0.58,0,0.82}
\definecolor{backcolour}{rgb}{0.95,0.95,0.92}
\definecolor{darkgreen}{rgb}{0.0, 0.6, 0.0}

\title{Fine-tuning in mixed Dark Matter models with Primordial Black Hole relics}

\author[1,2]{Amirah Aljazaeri \,\orcidlink{0009-0008-7032-8286},}
\author[1]{Christian T.~Byrnes\,\orcidlink{0000-0003-2583-6536},}

\affiliation[1]{Department of Physics and Astronomy, University of Sussex, Brighton BN1 9QH, UK\\}
\affiliation[2]{
Department of Physics, Taibah University, Madinah 42353, Saudi Arabia }

\emailAdd{aa2409@sussex.ac.uk}
\emailAdd{asjazaeri@taibahu.edu.sa}
\emailAdd{C.Byrnes@sussex.ac.uk}

\date{}

\abstract{We investigate the fine-tuning of a tripartite dark matter (DM) scenario involving ultra-light primordial black holes (PBHs), whose evaporation before Big Bang Nucleosynthesis leaves Planck-mass relics and produces DM particles together with an independently produced DM component. We uniformly evaluate the parameter sensitivity required to reproduce the observed DM abundance, $\Omega_{\rm DM}$. Considering thermal WIMP freeze-out, freeze-in, and QCD axion misalignment, and assuming PBHs form via collapse of perturbations following horizon entry, we find that the fine-tuning is normally dominated by the structure of the PBH relic abundance calculation rather than by the particle DM candidate or the details of PBH evaporation. In radiation-dominated cosmologies, the DM abundance is highly sensitive to the primordial curvature power spectrum because the PBH formation fraction depends exponentially on density fluctuations. Although an early PBH-dominated era dilutes pre-existing abundances and reduces the apparent tuning of the PBH abundance, inflationary fine-tuning remains required to produce the required initial large amplitude perturbations. We further examine alternative PBH formation channels, including supercooled first-order phase transitions and collapsing domain walls, and find that they replace the inflationary fine-tuning with alternative exponential sensitivities. 
We conclude that
a natural realization of an order unity PBH relic abundance is hard to motivate.
}

\begin{document}
\maketitle
\flushbottom
\section{Introduction}
\label{secintroduction}
The existence of dark matter (DM) is supported by a wide range of astrophysical and cosmological observations, including galactic rotation curves \cite{rubin1980rotational,rubin1970rotation}, gravitational lensing, e.g.~the Bullet Cluster \cite{clowe2006direct}, cosmic microwave background anisotropies \cite{aghanim2020planck}, and large-scale structure formation \cite{davis1985evolution,springel2005simulations}. Despite its overwhelming gravitational evidence, the microscopic nature of DM remains unknown \cite{bertone2005particle,bertone2018history}. Traditional single-component candidates, such as Weakly Interacting Massive Particles (WIMPs) \cite{goldberg1983constraint,ellis1984supersymmetric,jungman1996supersymmetric}, have long served as the canonical paradigm; however, increasingly stringent constraints from direct and indirect detection experiments \cite{aprile2018dark,aalbers2023first,ackermann2015searching} have motivated the exploration of more complex dark sectors involving multiple components and non-thermal production mechanisms \cite{feldman2010multicomponent,geng2015multi,mcdonald2002thermally,hall2010freeze}.

Primordial black holes (PBHs), initially proposed by Zel'dovich et al. \cite{zel1966hypothesis} and further developed by Hawking and collaborators \cite{hawking1971gravitationally,carr1974black}, provide a particularly compelling alternative to explain DM invoking neither new particles nor new physics \cite{chapline1975cosmological}. Unlike astrophysical black holes formed through stellar collapse, PBHs are theoretical black holes that could have formed in the early universe through the collapse of large density perturbations which may arise in inflationary models that produce pronounced peaks in the primordial power spectrum -- for detailed reviews, see \cite{carr2021constraints, sasaki2018primordial, arbey2024review, carr2026review}. 

During the radiation-dominated era, PBHs could form with a broad range of masses. Their initial mass is approximately equal to the horizon mass at the time the overdensity re-enters the horizon, 
\begin{equation}
    \label{eqMPBHi}
    M_{\rm PBH,i}
    \simeq
    M_{H} = 
    \frac{4\pi}{3}\rho_r
    \left(\frac{c}{H}\right)^3
    \simeq
    1\times10^{39}
    \left(\frac{t_{\rm i}}{\rm s}\right)
    \,\mathrm{g},
\end{equation}
where $H^{-1}$ corresponds to the Hubble radius at the time of formation, and $\rho_{\rm{r}}$ is the radiation energy density. Throughout this paper, for simplicity, we approximate the PBH mass distribution as nearly monochromatic and neglect the impact of accretion, as its effects are likely negligible on the small scales relevant to this work. 

A remarkable feature of sufficiently light PBHs is that they evaporate through Hawking radiation, thereby emitting all particle species lighter than the black hole temperature \cite{hawking1975particle,page1976particle}. Assuming a constant effective number of relativistic degrees of freedom for Hawking emission, $g_{*,H}(T_{\rm BH}) \simeq 108$, which is a good approximation for initial PBH masses $M_i \ll 10^{11}\,\mathrm{g}$ \cite{hooper2019dark}, the lifetime of an evaporating PBH is given by
\begin{equation}
\label{eqtev}
   t_{\rm{ev}} \simeq 4 \times 10^{-28}
    \left(\frac{M_{\rm{PBH,i}}}{\rm{g}}\right)^{3} \,\,\,\rm{s}.
\end{equation}

Therefore,
PBHs with initial masses below approximately $10^9\,\rm{g}$ evaporate before the onset of Big Bang Nucleosynthesis (BBN), making them especially relevant for early-universe cosmology and non-thermal DM production (while those with masses $M_{\rm{PBH,i}} \gtrsim 10^{15}$g have lifetimes exceeding the current age of the universe). Depending on the underlying quantum gravitational completion, evaporation may terminate at the Planck scale, leaving behind stable Planck-mass relics \cite{Carr1994,Chen2005,Green1997} which themselves behave as DM - see also our previous work \cite{Aljazaeri2025ftv}.

In this paper, we investigate a tripartite DM framework in which the total DM abundance is distributed among three distinct sectors: Planck-mass relics left behind after PBH evaporation, particles produced directly through Hawking radiation, and an independent DM component, originally taken to be thermal freeze-out WIMPs\footnote{We note that PBHs and WIMPs are incompatible only for sufficiently massive PBHs \cite{Lacki2010zf,Adamek2019gns,Carr2020mqm,Eroshenko2024dtb,Lavalle2026fhx}, unlike the ultra-light PBHs considered here.}. We then extend this framework by replacing the thermal freeze-out contribution with either freeze-in DM \cite{hall2010freeze,Bernal2017kxu} or QCD axions produced via the misalignment mechanism \cite{peccei1977cp,wilczek1978problem,preskill1983cosmology,abbott1983cosmological,marsh2016axion}, allowing us to investigate whether the degree of fine-tuning depends sensitively on the choice of DM production mechanism.

The primary objective of this paper is to quantify the naturalness of this tripartite framework. In particular, we evaluate the sensitivity of the total DM abundance to the fundamental parameters governing PBH formation, evaporation, and particle production. Our analysis focuses on the Barbieri--Giudice fine-tuning measure \cite{barbieri1988upper,ellis1986observables}, applied to quantities such as the primordial curvature power spectrum $P_\zeta$, the PBH mass $M_{\rm PBH}$, the DM particle parameters $m_{\chi}$, and their cross section $\langle \sigma v \rangle$. Since the PBH abundance depends exponentially on the amplitude of primordial fluctuations \cite{carr1975primordial,sasaki2018primordial,carr2021constraints}, even small changes in the inflationary sector can dramatically alter the resulting relic abundance \cite{sasaki2018primordial,byrnes2019steepest}, making PBHs a natural laboratory for studying cosmological fine-tuning \cite{cole2023primordial,Frolovsky:2023hqd,Stamou2024lqf,Lorenzoni2025kwn,Iovino2025tcv,profumo2026primordial,Ciaiolo:2026iss}. We quantify this sensitivity.  

A central aspect of this work is the comparison between different evaporation histories and cosmological backgrounds. We examine both scenarios in which PBHs evaporate during radiation domination and cases where they temporarily dominate the energy density of the universe before evaporating, leading to an early PBH-dominated era (ePBHd). In the latter case, entropy injection can dilute pre-existing species and generate attractor-like behaviour where the final DM abundance becomes largely independent of the initial PBH fraction \cite{carr1975primordial,macgibbon1987can,lennon2018black}. We additionally explore whether alternative PBH formation mechanisms, such as supercooled phase transitions \cite{caprini2016science,gouttenoire2024primordial} and domain wall collapse \cite{gouttenoire2024domainwall,gouttenoire2024domain}, alleviate or merely shift the underlying fine-tuning problem.

The structure of this paper is as follows. In Section~\ref{secTHEmodel}, we introduce the tripartite DM framework and derive the abundances of relics, evaporated particles, and thermal DM. Section~\ref{secFineTuning} quantifies the fine-tuning associated with the model in both radiation-dominated and early matter-dominated evaporation scenarios. Section~\ref{secAxions} extends the framework to axion DM and studies the robustness of the tuning under alternative particle candidates and non-standard cosmological histories. Finally, we investigate alternative PBH formation mechanisms and discuss whether they provide a more natural realization of PBH-driven dark sectors.

We adopt the notation and units summarized in \cref{tabnotation}. Throughout the paper, when referring to individual DM components, we use $\Omega_{X} \equiv f_{X} \,\Omega_{\rm DM}$. 
\begin{table}[ht]
\centering
\caption{Summary of the main quantities used in this work (D-less means dimensionless).}
\label{tabnotation}
\begin{tabular}{l p{8cm} p{2.5cm}  l}
\toprule
\textbf{Notation} & \textbf{Definition} & \textbf{Quantity} & \textbf{Unit} \\
\midrule
\textbf{Temperature}\\
\hline
$T_{\rm BH}$ & Hawking temperature & \cref{eqTbh} & K \\
$T_i$ & Background temperature at PBH formation & \cref{eqTi} & K \\
$T_{\rm evap}$ & Background temperature at the end of PBH evaporation, excluding heating from PBHs & \cref{eqTevap} & K \\
\hline
\textbf{Time}\\
\hline
$t_i$ & Time of PBH formation &  \cref{eqMPBHi} & s \\
$t_{\rm ev}$ & PBH evaporation time &  \cref{eqtev} & s \\
$t_{\rm fo}$ & WIMP freeze-out time &  \cref{eqtfo} & s \\
\hline
\textbf{d.o.f}\\
\hline
$g_*$ & Effective number of degrees of freedom of all particles radiated by the PBH & 7 - 108 & D-less \\
$g_s$ & Effective number of relativistic entropy degrees of freedom & 106.75 & D-less \\
$g_{\chi}$ & Internal degrees of freedom of the DM particle & 1 & D-less \\
$g_j$ & Internal degrees of freedom of the WIMP & $g_{\chi}$ &  D-less \\
$g_{{\rm a},H}$ & Effective Hawking degrees of freedom of a spin-0 QCD axion & $1.82$ &  D-less\\
\hline
\textbf{Abundance} \\
\hline
$\beta_i$ & Initial PBH abundance at the time of formation & \cref{eqBetai} & D-less \\
$\beta_{\rm c}$ & Critical initial PBH abundance separating radiation domination from an ePBHd era & \cref{eqBetac} & D-less \\
$f_{\rm relic}$ & DM fraction in leftover Planck-mass PBH relics & \cref{eqfrelic} & D-less \\
$f_{\chi}$ & DM fraction in particles produced via PBH evaporation & \cref{eqfchi} & D-less \\
$f_{\rm{WIMP}}$ & DM fraction in thermally produced WIMPs & \cref{eqOmegaWIMP} & D-less \\
$f_{\rm FI}$ & DM fraction via freeze-in & \cref{eqOmegaFI} & D-less \\
$f_{\rm axion}$ & DM fraction via QCD Axions & \cref{eqOmegaa} & D-less \\
\hline
\textbf{Others}\\
\hline
$\langle\sigma v\rangle$ & Thermally averaged annihilation cross section & $3\times10^{-26}$ & cm$^3$\,s$^{-1}$ \\
$\Delta N_{\rm eff}$ & Contribution of PBH-emitted dark radiation to the effective number of neutrino species & \cref{eqNeff} & D-less \\
$s_0$ & The present-day entropy density & $2.9\times10^9$ & $\text{m}^{-3}$ \\
\bottomrule
\end{tabular}
\end{table}


\section{The tripartite model - theoretical motivation and cosmic timeline}
\label{secTHEmodel}

While the $\Lambda$CDM paradigm successfully describes the large-scale structure of the universe, the microscopic nature of DM remains one of the most significant open questions in modern physics \cite{bertone2018history,aghanim2020planck}. 
In this section, we formalize a hybrid scenario where the observed DM abundance is distributed across three 
distinct constituents: leftover Planck-mass PBH relics  ($f_{\rm{relic}}$), DM particles produced via PBH evaporation ($f_{\rm{\chi}}$), and thermally produced WIMPs ($f_{\rm{WIMP}}$), where the total DM fraction is:
$$f_{\rm{DM}} = f_{\rm{relic}} + f_{\rm{\chi}} + f_{\rm{WIMP}} = 1.$$

This multi-component framework can alleviate the experimental constraints that challenge certain single-candidate DM models. In particular, the thermal freeze-out component need only account for a subdominant fraction of the observed DM abundance thereby reopening regions of parameter space that would otherwise be excluded by direct-detection limits. To understand the interplay between the various DM components, we examine both their physical properties and the chronology of their production in the early Universe. A summary of the components considered in this work is provided in Table~\ref{tabdmproperties}.
\begin{table}[ht]
\centering
\caption{Mass scales and physical origins of the DM components.}
\label{tabdmproperties}
\begin{tabular}{@{}lll@{}}
\toprule
\textbf{Component} & \textbf{Physical Origin} & \textbf{Characteristic Mass} \\ \midrule
\textbf{Relics} & Quantum Gravity Remnants & $M_{\rm Pl} \approx 2.18 \times 10^{-5}\,\text{g}$ \\
\textbf{Evaporated DM} & Hawking Spectrum & MeV $< m_{\chi} \ll M_{\rm{PBH}}$ \\
\textbf{Thermal WIMPs} & Thermal Freeze-out & $10$ GeV -- $1$ TeV \\ 
\textbf{QCD Axions} & Field oscillation & $10^{-6}$ eV -- $10^{-2}$ eV \\
\bottomrule
\end{tabular}
\end{table}   

\subsection{$f_{\rm relic}$ : PBH evaporation and relic formation}
The relic component originates from the final stages of PBH evaporation. Through Hawking radiation, a PBH continuously loses mass by emitting particles with a thermal spectrum. The corresponding Hawking temperature \cite{hawking1975particle}, $T_{\rm BH}$, is inversely proportional to the black hole mass and is given by 
\begin{equation}
\label{eqTbh}
T_{\rm BH} = \mathcal{G}_b
\frac{\hbar c^3}{8\pi G k_B M_{\rm PBH}}
=\mathcal{G}_b
\frac{c M_{\rm Pl}^2}{8\pi k_B M_{\rm PBH}} \,\,\, \text{K},
\end{equation}
where $\mathcal{G}_b$ denotes the graybody factor. As the PBH evaporates, its temperature increases, accelerating the emission process. While the semiclassical description predicts complete evaporation, quantum-gravitational effects are expected to become important as the PBH approaches the Planck scale. In several scenarios, these effects may halt the evaporation process and leave behind a stable remnant with mass $M_{\rm Pl}\simeq 2.18\times10^{-5}\,\mathrm{g}$. Such relics interact only gravitationally and therefore constitute viable cold DM candidates. In this work, following Ref.~\cite{Aljazaeri2025ftv}, we allow for the possibility that evaporation terminates at the Planck scale, leaving stable relics.

The present-day relic abundance, expressed as a fraction of the total DM density, $f_{\rm relic}\equiv{\Omega_{\rm relic}}/{\Omega_{\rm DM}}$, depends on both the initial PBH abundance, $\beta_i$, and the cosmological expansion history. For sufficiently small values of $\beta_i$, the Universe remains radiation dominated throughout the evaporation process. For larger initial abundances, however, the PBHs temporarily dominate the energy density before evaporating, leading to an early PBH-dominated era (ePBHd). Following Ref.~\cite{Aljazaeri2025ftv}, the relic fraction is then given by
\begin{equation}
\label{eqfrelic}
f_{\rm relic}
\simeq 2\left(1+\frac{\Omega_b}{\Omega_{\rm DM}}\right)
\left(\frac{M_{\rm Pl}}{M_{\rm PBH}}\right)
\times
\begin{cases}
\left(\dfrac{t_{\rm eq}}{t_i}\right)^{1/2}\beta_i,
& \beta_i \le \beta_c \quad (\mathrm{RD}),
\\[2mm]
\left(\dfrac{t_{\rm eq}}{t_{\rm ev}}\right)^{1/2},
& \beta_i > \beta_c \quad (\mathrm{ePBHd}),
\end{cases}
\end{equation}
where the transition between the two regimes occurs at the critical initial abundance
\begin{equation}
\label{eqBetac}
\beta_c =
\left(\frac{t_i}{t_{\rm ev}}\right)^{1/2}
\simeq 1.6\times10^{-6}
\left(\frac{M_{\rm PBH,i}}{1\,\mathrm{g}}\right)^{-1}.
\end{equation}

When satisfying this condition, the final relic abundance becomes independent of the initial PBH abundance, reflecting the loss of memory of the initial conditions. Consequently, the relic density is then determined primarily by the initial PBH mass rather than by $\beta_i$.

\subsection{$f_\chi$ : Particles generated by PBH evaporation as DM}
\label{secFchi}
PBH evaporation provides a non-thermal and approximately democratic production mechanism: any particle species with mass below the instantaneous Hawking temperature $T_{\rm BH}$ is emitted with a rate determined primarily by its number of degrees of freedom. As PBHs evaporate they populate both Standard Model and dark-sector states while injecting entropy into the radiation bath. The process ends when the PBH reaches the Planck scale, leaving a stable relic contribution, $f_{\rm relic}$.

To determine the DM abundance generated by PBH evaporation in an expanding universe, we define the initial PBH fraction at formation time $t_i$  as $ \beta_{i} \equiv \rho_{\text{PBH}}/\rho_{\text{rad}}$. If $N_{\chi}$ denotes the total number of particles $\chi$ emitted by a single PBH, the present-day yield (number-to-entropy density ratio) is 
\begin{equation*}
    \frac{n_{\chi}}{s} |_{\text{today}} = N_{\chi} \frac{n_{\rm PBH}}{s}|_{t_i} = N_{\chi} Y_{i}.
\end{equation*}
Using the definition of initial PBH abundance,
\begin{equation}
\label{eqBetai}
    \beta_{i} = \frac{\rho_{\rm{PBH,i}}}{\rho_{\rm{rad,i}}} = \frac{M_{\rm{PBH}} n_{\rm{PBH,i}}}{\rho_{\rm{rad,i}}} = M_{\rm{PBH}} \frac{s_{i} Y_{i}}{\rho_{\rm{rad,i}}} 
\end{equation}
where  $s_{i} = 2\pi^2  g_{*} T^3 /45$, $\rho_{\rm{rad,i}} = \pi^2 g_{*} T^4/30$, and assuming $g_* \simeq g_{*s}$ at high temperatures, we obtain:
\begin{equation}
\label{eqYi}
    {Y_i} = \frac{3}{4} \frac{T_i}{M_{\rm{PBH}}} \beta_i.
\end{equation}

Following \cite{baumann2007primordialblackholebaryogenesis,morrison2019melanopogenesis}, the total number of DM particles emitted by a single PBH can be estimated by dividing the energy radiated into the species $\chi$ by the average energy of a blackbody BH,
$$N_{\chi} \simeq \frac{g_{\chi}}{g_{*}} \int_{m_{\rm{f}}}^{m_{\rm{i}}}  \frac{dM}{3 T_{BH}}, $$
where $g_{\chi}$ and $g_*$ are the degrees of freedom of the DM species and of all radiated particles, respectively. 
We assume evaporation terminates at a Planck-mass relic, $m_{\rm f}=M_{\rm Pl}$.  Increasing the relic mass (e.g.~to $100M_{\rm Pl}$) primarily rescales the relic abundance while leaving the number of emitted DM particles essentially unchanged.  This slightly shifts the required initial PBH abundance and therefore the Press--Schechter fine-tuning, but the effect is small and does not qualitatively alter our conclusions.

Two regimes arise depending on the relation between $m_{\chi}$ and the initial Hawking temperature. For $m_{\chi}\leq T_{\rm BH}$, DM is emitted throughout the PBH lifetime. For $m_{\chi}>T_{\rm BH}$, emission becomes efficient only after the PBH has evaporated to a mass for which $T_{\rm BH}\simeq m_{\chi}$. The resulting particle multiplicity is
\begin{equation}
\label{eqNchi}
N_{\chi} \simeq \,4.2\, \frac{g_{\chi}}{g_{*}} \times
\begin{cases}
\left( \frac{M_{\rm{PBH}}}{M_{\rm Pl}} \right)^2 ,&  m_{\chi} \leq T_{\rm{BH}} \vspace {2mm} \\
\left( \frac{M_{\rm Pl}}{m_{\chi}} \right)^2, &  m_{\chi} > T_{\rm{BH}}.
\end{cases}
\end{equation}

In the heavy regime ($m_{\chi} > T_{\rm{BH}}$), emission is essentially blocked until the PBH evaporates down to a threshold mass of $m_i \simeq M_{\rm Pl}^2 / m_{\chi}$. Consequently, $N_{\chi}$ becomes independent of the initial mass $M_{\rm{PBH}}$ and is suppressed by the high mass of the DM particle.

The final DM yield depends on whether evaporation occurs during radiation domination or after the PBHs temporarily dominate the energy density. It is given by \cite{bernal2021dark}
\begin{equation}
\label{eqYchi}
    Y_{\chi} = \frac{3}{4} \frac{N_{\chi}}{M_{\rm{PBH}}} \times
    \begin{cases} 
    T_i \beta_i & \text{for } \beta_i \leq \beta_c \vspace{2mm} \\
    T_{\rm{evap}} & \text{for } \beta_i > \beta_c
    \end{cases}
\end{equation}
where $\beta_c$ is the critical initial PBH fraction required for PBHs to dominate the universe before evaporating. For $\beta_i<\beta_c$, the yield scales linearly with the initial abundance, whereas for $\beta_i>\beta_c$ the entropy released during PBH evaporation erases this dependence. 

The characteristic temperature scales for these regimes are determined by the initial PBH mass. The temperature at the time of formation, $T_i$, and the background temperature excluding heating from PBHs at the conclusion of evaporation, $T_{\rm{evap}}$, are given by the following scaling relations:
\begin{align}
    T_i &\simeq 10^{29} \left( \frac{M_{\rm{PBH}}}{1 \text{ g}} \right)^{-1/2} \text{ K,} \\
    T_{\rm{evap}} &\simeq 10^{23} \left( \frac{M_{\rm{PBH}}}{1 \text{ g}} \right)^{-3/2} \text{ K.}
\end{align}

The different mass scalings reflect the distinct expansion histories in the radiation-dominated and ePBHd regimes, see \cref{figCosmic} and Appendix~\ref{Appendix1}.

The corresponding DM fraction today from this process is
\begin{equation}
\label{eqfchi}
f_\chi \equiv \frac{\rho_\chi}{\rho_c\Omega_{\rm DM}}
=
\frac{m_{\chi} s_0}{\rho_c\,\Omega_{\rm DM}}
Y_\chi,
\end{equation}
where $\rho_c$ is the critical density and $s_0\simeq2.9\times10^9\,{\rm m}^{-3}$ is the present-day entropy density.

Finally, particles produced through Hawking evaporation are born highly relativistic, with characteristic energies of order $T_{\rm BH}$. Requiring this population to behave as cold (or sufficiently cool) DM imposes a lower bound on the particle mass. Following \cite{morrison2019melanopogenesis}, we adopt
\begin{equation*}
m_{\chi} \gtrsim 0.6~{\rm MeV}.
\end{equation*}

\subsubsection{Dark radiation generated by PBH evaporation}
\label{secDR}

The analysis of the previous section assumes that the particles emitted by PBHs become part of the DM abundance. However, if the emitted species are sufficiently light, they remain relativistic until late times and instead contribute to the dark-radiation (DR) component of the Universe. This is particularly relevant for particles such as QCD axions. In these cases, the constraint is set by the contribution of the emitted particles to the effective number of relativistic degrees of freedom, $\Delta N_{\rm eff}$, given by 
\begin{equation}
\label{eqNeff}
\Delta N_{\rm eff}
\simeq
13.71
\frac{g_{{\rm DR},H}}
     {g_{*,H}\, g_{*S}(T_{\rm RH})^{1/3}}
\times
\begin{cases}
1,
&
\beta_i \ge \beta_c,
\\[0.4cm]
\displaystyle
\beta_i
\left(
\frac{t_{\rm ev}}{t_i}
\right)^{1/2},
&
\beta_i < \beta_c,
\end{cases}
\end{equation}
where the two branches correspond to evaporation occurring after an ePBHd phase ($\beta_i \ge \beta_c$) and during radiation domination ($\beta_i < \beta_c$), respectively \cite{hooper2019dark}.

For the case of a single QCD axion emitted by ultra-light PBHs with masses $M_{\rm PBH}\lesssim10^6\,{\rm g}$, we obtain
\begin{equation*}
\Delta N_{\rm eff}
\simeq
0.049\,
\min\!\left[
1,\,
\beta_i
\left(
\frac{t_{\rm ev}}{t_i}
\right)^{1/2}
\right],
\end{equation*}
as derived in Appendix~\ref{Appendix4}. Comparing this prediction with the current observational limit, $\Delta N_{\rm eff}\lesssim0.2$ \cite{ACT2025tim, aghanim2020planck}, we find that dark-radiation constraints are weak for a single QCD axion. According to \cite{hooper2019dark}, they become relevant only in scenarios containing a large number of light hidden-sector degrees of freedom, such as axiverse-like models, where the contributions from many species can accumulate and significantly enhance $\Delta N_{\rm eff}$.

\subsubsection{The ratio of fractions: $f_\chi/f_{\rm{relic}}$}

To identify the boundary between relic- and evaporation-dominated dark sectors, we define the parity condition, $f_\chi = f_{\rm{relic}}$. This choice serves only as a convenient benchmark for separating the two regimes and does not affect the main conclusions of this work. The impact of adopting alternative ratios is discussed in Appendix~\ref{Appendix2}. Using the expressions derived in the previous sections, this condition determines the locus in the $(m_{\chi}, M_{\rm{PBH}})$ plane where the DM abundance produced through Hawking evaporation equals that stored in Planck-mass remnants. 

\begin{figure}[ht]
    \centering
    \includegraphics[width=1\linewidth]{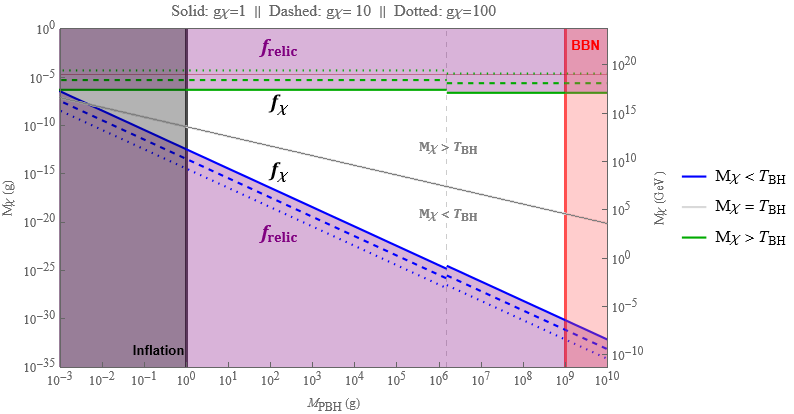}
    \caption[...]{
    The parity condition $f_{\chi} = f_{\rm relic}$ is shown in the $(M_{\rm PBH}, m_{\chi})$ plane. The blue curves correspond to the light DM regime ($m_{\chi} < T_{\rm BH}$), while the green curves represent the kinematically suppressed heavy regime ($m_{\chi} > T_{\rm BH}$). Different line styles (solid, dashed, dotted) indicate dark sector degrees of freedom $g_{\chi} \in \{1, 10, 100\}$. The gray  and red shaded regions denote constraints from inflation and BBN, respectively. The discontinuity near $M_{\rm PBH} \sim 10^6~\mathrm{g}$ marks the transition from radiation domination to an ePBHd era. Regions where $f_{\chi} > f_{\rm relic}$ are unshaded, whereas the shaded purple region corresponds to $f_{\chi} < f_{\rm relic}$. The results are independent of the initial PBH fraction $\beta_i$. Increasing $g_{\chi}$ leads to exclusion by the Planck-mass relic bound.}
    \label{figfxfr}
\end{figure}

For light DM, ($m_{\chi} < T_{\rm{BH}}$), the parity condition is
\begin{equation}
    m_{\chi} \approx M_{\rm Pl} \left( \frac{g}{g_{\chi}} \right) \left( \frac{M_{\rm Pl}}{M_{\rm{PBH}}} \right)^2,
\end{equation}
yielding the characteristic slope visible in the blue curves of \cref{figfxfr}. For heavy DM, $m_{\chi} > T_{\rm BH}$, particle emission is suppressed until the Hawking temperature reaches the mass threshold. The parity condition then becomes approximately independent of the initial PBH mass,
$m_{\chi} \sim M_{\rm Pl}(g_\chi/g)$,
corresponding to the horizontal green branches in \cref{figfxfr}. However, these scales are generally considered nonphysical, as they exceed standard DM mass expectations and lie beyond the regime tof Hawking evaporation, approaching the Planck scale. 

The transition between the radiation dominated and ePBHd regimes introduces a shift in the parity curves associated with the boundary $\beta_i = \beta_c$. As illustrated in \cref{figfxfr}, the qualitative behavior of the parity lines in the ePBHd regime does not deviate significantly from the radiation dominated case. We therefore restrict our analysis to the radiation-dominated regime. Furthermore, we assume $g_{\chi}=1$ throughout the remainder of this work as a representative benchmark. 

\subsection{$f_{\rm{WIMP}}$: Thermal WIMP}
\label{secWIMPs}
Complementing the non-thermal DM components produced by PBH evaporation and PBH relics, we consider a thermal WIMP component generated through the standard freeze-out mechanism. As the Universe cools, WIMPs initially in thermal equilibrium with the primordial plasma undergo freeze-out once their annihilation rate falls below the Hubble expansion rate. Their relic abundance is determined by the thermally averaged annihilation cross section $\langle \sigma v \rangle$ and the WIMP mass $m_{\chi}$, establishing an initial fraction $f_{\rm{WIMP}}$ that may subsequently be diluted by entropy injection from PBH evaporation. In our benchmark scenarios, we typically assign the WIMP component either a subdominant or dominant role (e.g., $1\%$ or $99\%$ of $\Omega_{\rm DM}$) to investigate how the PBH-produced components account for the remaining DM abundance. Consequently, in the ePBHd scenario, the entropy released during PBH evaporation dilutes the pre-existing WIMP population, requiring a significantly larger annihilation cross section than in the standard radiation-dominated cosmology to reproduce the observed DM density.

In this work, we focus primarily on the freeze-out (FO) production of WIMP DM. However, DM may also be produced through the freeze-in (FI) mechanism. In contrast to freeze-out, where DM is initially in thermal equilibrium with the Standard Model plasma before decoupling, freeze-in occurs when DM interactions are so feeble that thermal equilibrium is never established. Instead, the DM abundance is gradually built up through rare collisions or decays of particles in the thermal bath. As can be inferred from the results of Ref.~\cite{cheek2022primordial}, including a freeze-in component does not qualitatively alter our main conclusions. The key difference is that freeze-in DM never thermalizes with the products of PBH evaporation, allowing all three components of our tripartite DM scenario to contribute simultaneously to the total DM abundance. In \cref{secFI}, we extend our analysis to this scenario and quantify the corresponding fine-tuning of the model in the freeze-in case.

The viability of the tripartite DM scenario depends crucially on the relative timing of WIMP freeze-out and PBH evaporation. 
If PBHs evaporate while WIMPs are in thermal equilibrium with the primordial plasma, the evaporated DM particles thermalize and lose their non-thermal origin, effectively reducing the dark sector to fewer independent components. Conversely, if PBHs evaporate after WIMP freeze-out, the thermal and non-thermal populations remain distinct, allowing all three DM components to coexist.
We therefore determine the freeze-out time, $t_{\rm fo}$, and compare it with the PBH evaporation time, $t_{\rm ev}$.

Assuming freeze-out occurs during the radiation-dominated era, the energy density is dominated by relativistic species,
\begin{equation*}
\rho_r=\frac{\pi^2}{30}g_*T^4
\left(\frac{k_B^4}{\hbar^3c^5}\right) \rm g \,m^{-3},
\qquad
H^2=\frac{8\pi G}{3}\rho_r,
\end{equation*}
with $H=1/(2t)$. Solving for cosmic time yields
\begin{equation}
\label{eqtfo}
    t_{\rm{fo}} =  \frac{1}{2}\sqrt{\frac{90}{8\pi^3 g_*}} \frac{M_{\rm{pl}}}{T_{\rm{fo}}^2} \approx 3.15 \times 10^{19} \left( \frac{T_{\rm{fo}}}{\rm{K}}\right) ^{-2} \text{s}, 
\end{equation}
where
$g_*$ is the effective number of relativistic degrees of freedom. The freeze-out temperature is given by $T_{\rm fo}=m_{\chi}/x_{\rm fo}$, where $x_{\rm fo}$ is a dimensionless parameter that will be defined later and whose value depends on the cosmic history.

\begin{figure}[ht]
    \centering
    \includegraphics[width=1\linewidth]{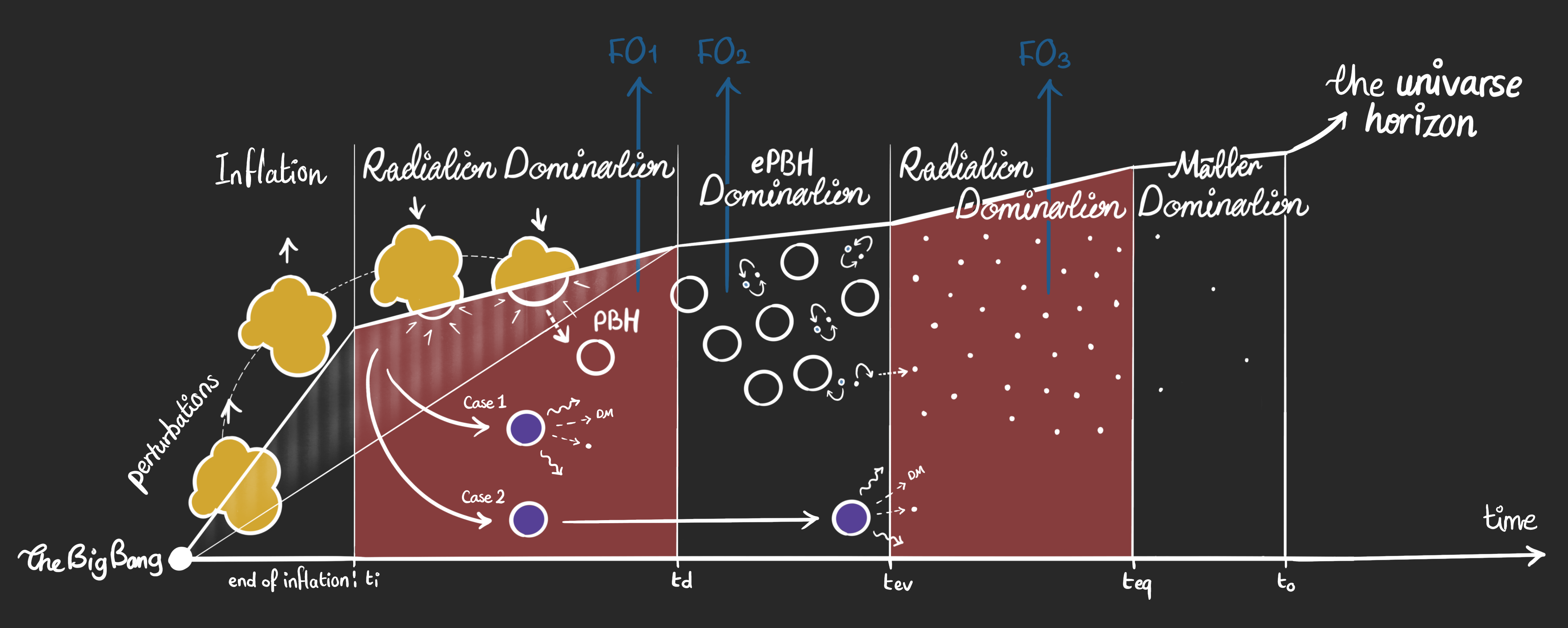}
    \caption[...]{Schematic cosmic history illustrating PBH formation and evolution from the end of inflation to the present epoch. The markers $FO_1$, $FO_2$, and $FO_3$ indicate the different WIMP freeze-out epochs considered in this work. If PBHs are sufficiently abundant they dominate before decaying as shown by case 2 (making $FO_2$ possible), else case 1 follows without PBH domination.}
    \label{figCosmic}
\end{figure}

Below a critical PBH mass, evaporation occurs prior to WIMP freeze-out ($t_{\rm ev}<t_{\rm fo}$). In this case, the particles emitted by PBH evaporation thermalize with the primordial plasma, preventing them from remaining a distinct non-thermal DM component and effectively reducing the dark sector to a bipartite scenario. Above this critical mass, the ordering is reversed ($t_{\rm ev}>t_{\rm fo}$). The thermal WIMP population freezes out before PBH evaporation, allowing the non-thermal $f_\chi$ and relic components to survive as independent DM constituents. Whether this occurs in a radiation-dominated or ePBHd Universe further determines the amount of entropy injection experienced by the thermal WIMP population. Consequently, the cosmological evolution is naturally divided into two main cases, each containing several sub-scenarios, as illustrated in \cref{figCosmic}. 

\textbf{Case 1: PBHs never dominate the cosmic energy density}
\begin{itemize}
    \item \textbf{If $t_{\rm fo}<t_{\rm ev}$:} WIMPs freeze out before PBH evaporation. The PBHs subsequently evaporate into a Universe that already contains a thermal WIMP population.

    \item \textbf{If $t_{\rm fo}>t_{\rm ev}$:} PBHs evaporate while WIMPs are still in thermal equilibrium. Consequently, the evaporated particles thermalize with the plasma and may become part of the thermal WIMP population.
\end{itemize}
For $M_{\rm PBH}\lesssim10^6\,\mathrm{g}$ and WIMP masses up to approximately $1\,\mathrm{TeV}$, we find that PBH evaporation precedes WIMP freeze-out. We therefore identify $M_{\rm PBH}\simeq10^6\,\mathrm{g}$ as the critical PBH mass separating the bipartite and tripartite DM regimes.

For Case 1, following \cite{bernal2022wimps}, the dimensionless freeze-out parameter $x_{\rm fo}$ is determined analytically using the $W_{-1}$ branch of the Lambert $W$ function, avoiding the usual iterative solution:
\begin{equation}
x_{\rm fo} = -\frac{1}{2} W_{-1} \left[ -\frac{8\pi^5}{45} \frac{g_s}{g_j^2} \frac{1}{(M_{\rm Pl} m_{\chi} \langle\sigma v\rangle)^2} \right],
\end{equation}
where $g_j$ denotes the internal degrees of freedom of the WIMP and $g_s$ is the effective number of relativistic entropy degrees of freedom.

Using the resulting value of $x_{\rm fo}$, the present-day WIMP relic abundance, $\Omega_{\rm WIMP}=f_{\rm{WIMP}}\,\Omega_{\rm DM}$, is calculated as
\begin{equation}
\label{eqOmegaWIMP}
\Omega_{\rm WIMP}^{\rm sc}
=
\frac{s_0}{\rho_c}
\left(
\frac{15}{2\pi(g_s+g_j)}
\sqrt{\frac{g_s}{10}}
\frac{1}{M_{\rm Pl}}
\right)
\frac{x_{\rm fo}}{\langle\sigma v\rangle}.
\end{equation}

The above expression describes the standard radiation-dominated cosmology (SC). In the presence of an ePBHd era, the WIMP abundance is determined by the ordering of the freeze-out and PBH evaporation times, resulting in three possible scenarios:
\smallskip

\textbf{Case (2): When PBHs dominate}

\begin{itemize}
    \item \textbf{$FO_1$: WIMPs freeze out before ePBHd}

   In this case, the tripartite DM scenario remains valid, but the WIMP abundance is diluted by the subsequent ePBHd era:
$$\Omega_{\rm WIMP}
= \Omega_{\rm WIMP}^{\rm sc}
\left(\frac{a_{\rm d}}{a_0}\right)^3,$$
where $a_{\rm d}$ denotes the scale factor at the onset of the ePBHd era, and $a_0$ is the present-day scale factor.
    \item \textbf{$FO_2$: WIMPs freeze out during ePBHd}

    This case requires a more detailed treatment of the freeze-out process in a non-standard cosmological background, as discussed in Ref.~\cite{cheek2022primordial}. However, such a treatment is beyond the scope of this work and has a negligible impact on our final results. Therefore, we approximate this scenario using the same result as in the $FO_1$ case.

    \item \textbf{$FO_3$: WIMPs freeze out after ePBHd}

    In this case, the Universe has returned to the standard radiation-dominated phase before WIMP freeze-out, and therefore
    \[
    \Omega_{\rm WIMP}
    =
    \Omega_{\rm WIMP}^{\rm sc}.
    \]
    Since the PBH-produced particles thermalize before WIMP freeze-out, the tripartite DM scenario is reduced to a bipartite one.
\end{itemize}

\section{The Fine-Tuning Definition and Measure of the Model}
\label{secFineTuning}

In this section, we quantify the degree of fine-tuning required to reproduce the observed DM abundance, $\Omega_{\rm DM}h^2 \simeq 0.12$, within the tripartite DM scenario. The total DM abundance is given by the sum of the thermal WIMP, PBH relic, and PBH evaporation contributions, 
\begin{equation*}
\Omega_{\rm DM}
=
\Omega_{\rm WIMP}
+
\Omega_{\rm relic}
+
\Omega_{\chi}.
\end{equation*}

To quantify the sensitivity of the predicted DM abundance to the model parameters, we employ the Barbieri--Giudice (BG) fine-tuning measure \cite{barbieri1988upper},
\begin{equation*}
\Delta_p =
\left|
\frac{\partial\ln\Omega}{\partial\ln p}
\right|
=
\left|
\frac{p}{\Omega}
\frac{\partial\Omega}{\partial p}
\right|,
\end{equation*}
where $p$ denotes an input parameter of the model. A large value of $\Delta_p$ indicates that a small fractional variation in $p$ produces a much larger fractional change in the predicted DM abundance.

The BG measure has two well-known limitations. First, it is a local sensitivity measure, probing only the response of the model to infinitesimal parameter variations around a chosen benchmark point. It therefore does not address broader notions of naturalness, such as the size of the viable parameter space or the stability of the model under radiative corrections. Secondly, it is not reparameterization invariant, meaning that the inferred degree of fine-tuning depends on the choice of fundamental parameters. These limitations have been discussed extensively in the literature and reflect the fact that different definitions of naturalness quantify different physical concepts \cite{profumo2026primordial,Profumo:2026kfy}. We focus on the local BG measure, since our objective is to identify the parameters to which the DM abundance is most sensitive.

A key parameter in our analysis is the primordial curvature power spectrum, $P_{\zeta}$ (or equivalently its peak amplitude $A_s$), which determines the initial abundance of PBHs. Assuming Gaussian primordial density fluctuations, the Press--Schechter formalism \cite{PressSchechter1974,carr1975primordial} relates the initial PBH mass fraction to the curvature power spectrum through
\begin{equation*}
\beta(P_{\zeta})
=
\operatorname{Erfc}
\left(
\frac{\delta_c}{\sqrt{2P_{\zeta}}}
\right),
\end{equation*}
where we adopt the collapse threshold $\delta_c\simeq0.45$ \cite{musco2005computations}. Because $\beta$ depends exponentially on $P_{\zeta}$, even modest fine-tuning in the initial PBH abundance corresponds to a much stronger sensitivity to the primordial power spectrum.

Finally, the fine-tuning measure $\Delta_{P_{\zeta}}$ obtained in this work should be regarded as a lower bound on the total tuning of the complete model. Here, $P_{\zeta}$ is treated as an independent parameter, while in reality it is determined by the parameters of the inflationary potential, $V(\phi)$. Consequently, the sensitivity to the fundamental parameters of the inflationary Lagrangian is expected to exceed that quantified by $\Delta_{P_{\zeta}}$. Indeed, Ref.~\cite{cole2023primordial} quantified the significant fine-tuning in single-field inflationary models capable of generating the required enhancement of the primordial power spectrum (from $10^2-10^8$), suggesting that the overall tuning of a complete inflationary PBH scenario is orders of magnitude larger than the values reported here.

\subsection{Case 1: Evaporation during radiation domination}
\label{secFineRD}
We first consider the radiation-dominated (RD) scenario, corresponding to $\beta_i<\beta_c$, in which PBHs evaporate before dominating the energy density of the Universe. In principle, the total DM abundance consists of all three components of the model. However, as discussed in \cref{secWIMPs},  for this mass range the relevant DM abundance reduces to 
$$ 
\Omega_{\rm total}
=\Omega_{\rm relic}+\Omega_{\rm WIMP}.
$$

Moreover, obtaining an appreciable PBH relic abundance requires relatively large values of $\beta_i$. For initial PBH masses above $M_{\rm PBH}\simeq10^6\,{\rm g}$, relics cannot constitute a significant fraction of the DM, even if PBHs temporarily dominate the Universe, due to their late evaporation time.

The corresponding Barbieri--Giudice fine-tuning measures are defined as
\begin{equation}
\label{eqFT}
\Delta_p
=
\left|
\frac{\partial\ln\Omega_{\rm total}}
{\partial\ln p}
\right|,
\qquad
p\in
\left\{
P_{\zeta},
M_{\rm PBH},
m_{\chi},
\langle\sigma v\rangle
\right\}.
\end{equation}

\begin{figure}[ht]
    \centering
    \includegraphics[width=0.9\linewidth]{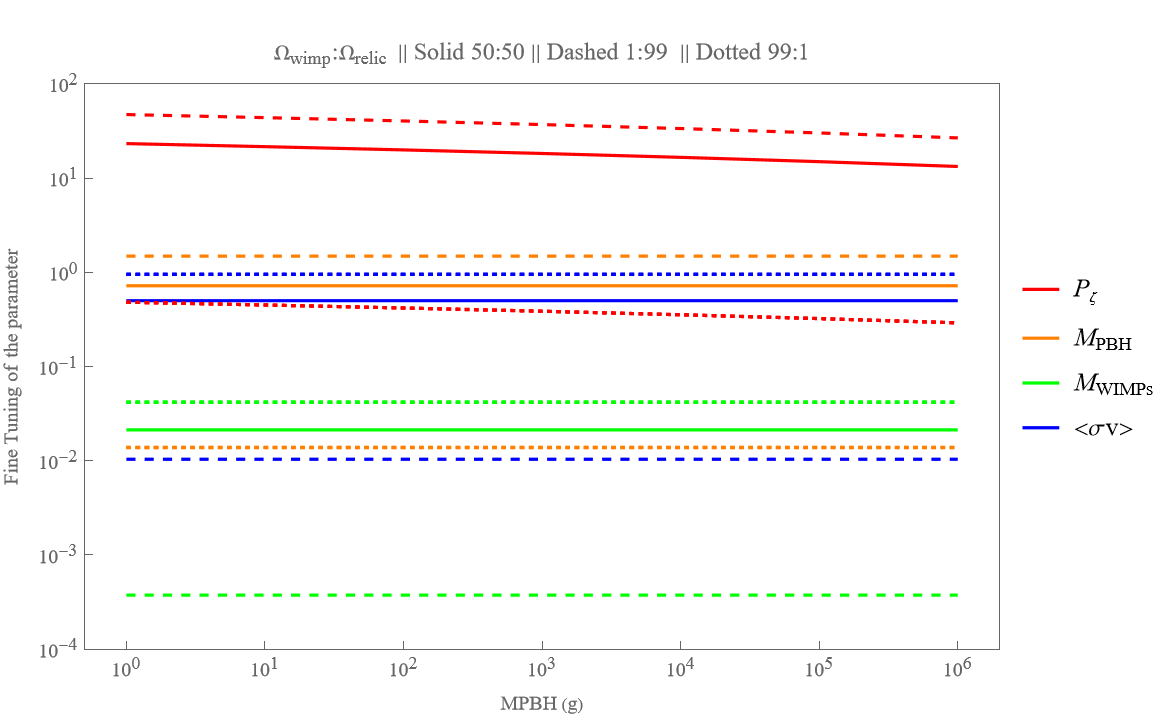}
    \caption{Barbieri--Giudice fine-tuning measures assuming no early PBH domination, where $\Omega_{\rm total}=\Omega_{\rm relic}+\Omega_{\rm WIMP}$. The sensitivities to the primordial curvature power spectrum $P_{\zeta}$ (red), the initial PBH mass $M_{\rm PBH}$ (orange), the WIMP mass $M_{\rm WIMP}$ (green), and the annihilation cross section $\langle\sigma v\rangle$ (blue) are shown, using $M_{\rm WIMP}=m_{\chi}=100~{\rm GeV}$. $P_{\zeta}$ dominates the dominant fine-tuning unless the relic contribution is negligible.}
    \label{figFineRD}
\end{figure}

For the numerical analysis shown in \cref{figFineRD}, we adopt the representative benchmark
$m_{\chi}=M_{\rm WIMP}=100~{\rm GeV}$. The resulting fine-tuning measures exhibit a clear hierarchy in parameter sensitivity unless there is no fine-tuning as is the case with a negligible relic contribution. If PBH relics dominate, or there is a comparable contribution from both components, the primordial curvature power spectrum remains the dominant source of fine-tuning, with $\Delta_{P_{\zeta}}\sim10$--$100$. Notably, all fine-tuning measures except $\Delta_{P_{\zeta}}$ (red) are independent of the initial PBH mass.

\subsection{Case 2 : Evaporation after early PBH domination}
\label{secFineMD}
The transition to an early matter-dominated era ($\beta_i > \beta_c$) fundamentally alters the fine-tuning landscape and offers a unique ``attractor" behavior. 
In this regime, the substantial entropy injection from PBH evaporation dilutes all pre-existing species, effectively ``resetting" the yields. As shown in \cref{secFchi}, the final DM yields $Y_{\chi}$ become largely independent of the initial fraction $\beta_i$ due to this. 

\begin{figure}[ht]
    \centering
    \includegraphics[width=0.9\linewidth]{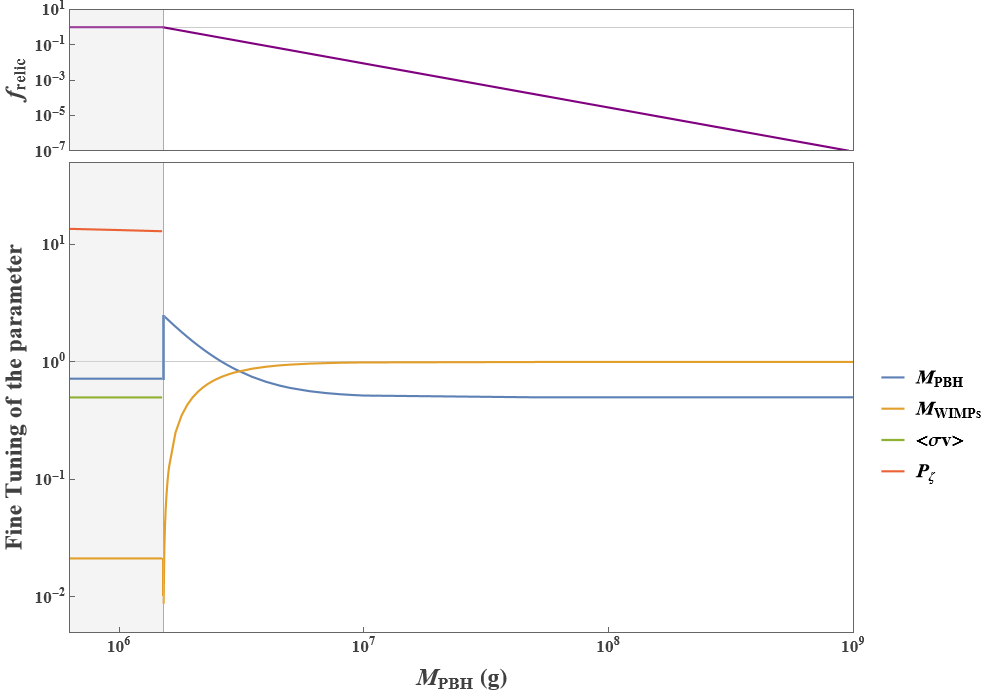}
   \caption[...]{\textbf{Bottom plot:} Barbieri--Giudice fine-tuning measures for the early PBH-dominated scenario. The sensitivities to the primordial curvature power spectrum $P_{\zeta}$ (red) which is zero where not shown, the initial PBH mass $M_{\rm PBH}$ (blue), the WIMP mass $M_{\rm WIMP}$ (yellow), and the annihilation cross section $\langle\sigma v\rangle$ (green) are shown as functions of the initial PBH mass. The gray vertical line marks the transition to the ePBHd regime. Before this transition, all four parameters contribute to the total fine-tuning. Beyond it, the thermal WIMP abundance is assumed to be erased by the ePBHd era, leaving the PBH relic and PBH-produced DM components to account for the observed DM abundance. Consequently, the sensitivity to $\langle\sigma v\rangle$ becomes negligible, the sensitivity to $M_{\rm WIMP}$ is replaced by that of the PBH-produced DM mass $m_{\chi}$, and the Press--Schechter relation between $P_{\zeta}$ and $\beta$ used in the RD case is no longer applicable.
   \textbf{Top plot:} The fraction of DM in PBH relics, where the value in the shaded region is the maximum allowed by observations but on the right it is the theoretical maximum (reached via $\beta>\beta_c$).}
    \label{figOmegaM}
\end{figure}

Therefore, in both $FO_1$ and $FO_2$, $\Omega_{\rm{WIMP}}$ is washed by ePBHd, and $\Omega_{\rm{total}}$ became $ \Omega_{\rm{relic}} + \Omega_{\chi}$. We examine this scenario where the total DM abundance is comprised of both relics and evaporated particles, as shown in \cref{figOmegaM}, to naively conclude that the model is no longer fine-tuned. 

Nevertheless, despite the ``naturalness" of the $\beta$-independence in the ePBHd phase, the underlying power spectrum amplitude must still be enhanced by many orders of magnitude to ensure the universe enters the MD phase in the first place, maintaining the overarching theme of high sensitivity in the primordial era. This is a good example of how different fine-tuning criteria are inequivelant and an example of where a simple application of the Barbieri-Giudice criteria leads to a misleading conclusion, since it is certainly false to conclude from $\Delta=0$ that there is no fine-tuning with respect to the power spectrum. Viewed on a log scale, for the vast majority of initial amplitudes between $10^{-9}$ and unity no PBH domination (or even formation) will occur, but once the amplitude is $\gtrsim10^{-2}$ a local attractor is reached and hence $\Delta=0$ with respect to this parameter. This is a limitation of measuring fine-tuning only locally, and it does not say anything about what fraction of the parameter prior volume leads to an `interesting' amplitude of $f_{\rm PBH}$ \cite{profumo2026primordial,Profumo:2026kfy}. 

This brings in a potential use of the `Wilson' fine-tuning criterion promoted in the PBH context in \cite{Iovino2025tcv}, who asked whether the value of $\Delta_{P_{\zeta}}$ at the fiducial point is similar to the value at nearby points, interpreting an affirmative answer as natural even if $\Delta_{P_{\zeta}}\gg1$, and concluding that all of the single-field inflationary models they studied were natural.\footnote{We note that the answer is sensitive to the chosen parameter range. Iovino and Riotto allow the amplitude to vary by about a factor of 2, which corresponds to $\sim$8 orders of magnitude range in $f_{\rm PBH}$, but if the power spectrum amplitude was allowed to vary from the CMB amplitude of $10^{-9}$ then their parameter would also be much larger than unity.} In the PBH dominated era case the tuning parameter they use would also be zero, rather than the expected order unity for `non-fine-tuned' cases. It's unclear how this should be interpreted, but this could be an interesting application of a test which compares fine-tuning in neighbouring parts of parameter space, since $\Delta_{P_{\zeta}}$ has a discontinuity at $\beta=\beta_c$.

\subsection{Freeze-In (FI) feebly interacting massive particles in the tripartite model}
\label{secFI}
We consider the standard infrared (IR)-dominated freeze-in mechanism during radiation domination, following Ref.~\cite{hall2010freeze} (a simplified expression is in \cite{profumo2026primordial}), within the framework of ultra-light PBH masses. Assuming a mediator of mass $M_D$ with decay coupling $y_D$ to DM, and a reheating temperature satisfying $T_{\rm RH}\gg M_D$, the present-day relic abundance produced via freeze-in is given by
\begin{equation}
\label{eqOmegaFI}
\Omega_{\rm FI}
=
\frac{s_0 m_{\chi}}{\rho_c}
\left(
\frac{45\,\chi_{\rm FI}}{\pi^{4}g_{*}^{3/2}}
\right)
\frac{y_D^{2}M_{\rm Pl}}{M_D},
\end{equation}
where $\chi_{\rm FI}$ is a dimensionless phase-space integral of order unity that depends weakly on the spin of the mediator, and we adopt $g_* = 100$.

Unlike freeze-out, the DM particles never attain thermal equilibrium with the Standard Model plasma due to their extremely feeble interactions. Instead, the DM abundance is gradually built up through the decays of the thermal bath mediator $D$ into pairs of DM particles. Therefore, our tripartite model still holds even if FI-produced DM forms after PBH evaporation.

As a benchmark, we fix the mediator mass to $M_D = 100~{\rm GeV}$. For the tripartite DM scenario, in which freeze-in contributes $34\%$ of the total DM abundance while the remaining $66\%$ is equally shared between the other two production mechanisms, the corresponding Barbieri--Giudice fine-tuning measures are
\begin{equation*}
\Delta_{y_D} \simeq 0.68,
\qquad
\Delta_{m_{\chi}} \simeq 0.67,
\end{equation*}
which are independent of the PBH mass over the range $1~{\rm g} \leq M_{\rm PBH} \leq 10^{6}~{\rm g}$.

Consequently, the dominant contribution to the overall fine-tuning continues to arise from the primordial curvature power spectrum, with
\begin{equation*}
\Delta_{P_{\zeta}} \simeq 19-32 ,
\end{equation*}
over the same PBH mass range.

\section{QCD axion}
\label{secAxions}
We now consider the QCD axion as an alternative to the thermal DM component in our tripartite model. The QCD axion is a non-thermal hypothetical pseudo-Nambu–Goldstone boson that arises from the spontaneous breaking of the Peccei–Quinn symmetry, proposed to solve the strong CP problem in quantum chromodynamics \cite{peccei1977cp,weinberg1978new}. After the QCD phase transition, the axion develops a small, temperature-dependent mass through non-perturbative QCD effects. Initially displaced from the minimum of its potential, the axion field remains frozen until the Hubble expansion rate falls below the axion mass, after which it begins coherent oscillations via the misalignment mechanism. These oscillations behave as a condensate of cold, non-relativistic particles, allowing axions to contribute to the present-day DM relic abundance \cite{di2020landscape}. In the following, we summarise the dependence of the axion relic abundance, $\Omega_a$, on different possible cosmic histories. 

\subsection{Standard cosmology (SC) scenarios}
The axion relic abundance in standard cosmology depends primarily on the initial misalignment angle $\theta_i$ and the axion mass $m_a$. The allowed range of $\theta_i$, and therefore the resulting axion abundance, is determined by the
cosmological history of the PQ symmetry breaking relative to inflation. This leads to two distinct scenarios: 
\begin{itemize}
    \item \textbf{Post-inflationary:} If the PQ symmetry is broken after inflation, the observable Universe would contain many patches with different values of $\theta_i$, and the axion abundance receives contributions from both the misalignment mechanism and the decay of topological defects, such as axion strings and domain walls. In this case, the initial misalignment angle is averaged statistically over the possible field values, giving an effective value $\theta_i = \pi/\sqrt{3} \approx 1.81$.  
    
    \item \textbf{Pre-inflationary:} If the PQ symmetry is broken before or during inflation, inflation homogenizes the axion field and dilutes any topological defects. The initial misalignment angle is then a single free parameter across the observable Universe, allowing a wider range of axion masses by adjusting $\theta_i$ to achieve the required relic abundance.
\end{itemize}
A summary of these two scenarios is provided in Table~\ref{tab:axion_scenarios} and \cite{venegas2021relic}. For large initial angles, $\theta_i \rightarrow \pi$, the harmonic approximation to the axion potential becomes inaccurate. In this regime, anharmonic effects delay the onset of oscillations as the axion field remains near the potential maximum for longer. This results in an enhanced relic abundance, which can be accounted for by an anharmonic correction factor $F(\theta_i)$; see Ref.~\cite{di2020landscape, venegas2021relic}.
\begin{table}[h!]
    \centering
    \caption{Comparison of Axion Cosmological Scenarios}
    \label{tab:axion_scenarios}
    \begin{tabular}{lll}
        \toprule
        \textbf{Feature} & \textbf{Post-Inflationary} & \textbf{Pre-Inflationary} \\ \midrule
        \textbf{PQ Breaking} & After inflation & Before/During inflation \\
        \textbf{Mechanisms} & Misalignment/String/Wall Decays & Only Misalignment \\
        \textbf{Axion Field} & Inhomogeneous (multi-patch) & Homogeneous (single patch) \\
        \textbf{Initial Angle ($\theta_i$)} & Averaged: $\pi/\sqrt{3} \approx 1.81$ & Free parameter ($-\pi < \theta_i < \pi$) \\
        \textbf{Mass Range ($m_a$)} & $5 \times 10^{-6}$ eV to $10^{-2}$ eV & Flexible \\ \bottomrule
    \end{tabular}
\end{table}

For a standard cosmological expansion history, we adopt the following approximation for the axion relic abundance as a function of the zero-temperature axion mass, $m_a$, \cite{venegas2021relic}:
\begin{equation}
\label{eqOmegaa}
\Omega_a^{sc} \approx
\begin{cases} 
0.17\,\theta_i^2
\left( \frac{m_a}{5.6\,\mu{\rm eV}} \right)^{-7/6},
& m_a \gtrsim 10^{-5}\,\mu{\rm eV}, \\[6pt]
0.006\,\theta_i^2
\left( \frac{m_a}{5.6\,\mu{\rm eV}} \right)^{-3/2},
& m_a \lesssim 10^{-5}\,\mu{\rm eV}.
\end{cases}
\end{equation}

In the post-inflationary scenario, the requirement that axions reproduce the observed DM abundance, $\Omega_a \simeq \Omega_{\rm DM}\approx0.26$, constrains the axion mass to the classic window $5\times10^{-6}\,{\rm eV}\lesssim m_a \lesssim10^{-2}\,{\rm eV}$.

\subsubsection{The fine tuning of the tripartite model}
To investigate whether the fine-tuning of the tripartite framework depends on the nature of the particle DM candidate, we repeat the fine-tuning analysis for the QCD axion. The total DM abundance is then expressed as 
$$f_{\rm{DM}} = f_{\rm{relic}} + f_{\rm{\chi}} + f_{\rm axion} = 1,$$
where the axion contribution is given by $f_{\rm axion} = \Omega_a ^{sc} / \Omega_{\rm{DM}}$, assuming axion production through the misalignment mechanism with the
post-inflationary initial angle $\theta_i = \pi/\sqrt{3}$. 

We found very similar results, where the sensitivity of the model remains dominated by the power spectrum ($P_{\zeta}$) parameter. This leads to a distribution similar to that shown in \cref{figFineRD}, suggesting that the tripartite balance is robust across different particle candidates.

We find that the resulting fine-tuning behaviour is very similar to that of the thermal DM case. In particular, the sensitivity remains dominated by the primordial power spectrum amplitude $P_{\zeta}$, with the fine-tuning distribution closely following that shown in \cref{figFineRD}. The small differences arise from the treatment of the PBH evaporation-produced component: in the thermal DM case, the emitted particles thermalize and contribute to the DM abundance as thermal DM, whereas in the QCD axion case the corresponding relativistic contribution is treated as dark radiation, as discussed in \cref{secDR}. This indicates that the tripartite balance is largely insensitive to the identity of the non-PBH DM component and remains robust across different DM candidates.

\subsection{Non-standard cosmology (NSC) scenarios}
We next investigate whether a non-standard cosmological history, instead of the standard radiation-dominated evolution, modifies the fine-tuning of the model. Such scenarios, including early matter domination and kination, alter the expansion history of the Universe and can introduce entropy dilution of the axion abundance. This significantly modifies the relation between the axion mass and its relic abundance, allowing a wider range of axion masses than the standard ``axion window'' permits.

Nevertheless, within the scope of this study, these modifications change the overall fine-tuning by no more than $\mathcal{O}(1)$. This is because the axion relic abundance retains a smooth power-law dependence on the relevant model parameters, such as the early-universe temperature scales ($T_{\rm end}$) and the axion mass ($m_a$), see Appendix~\ref{Appendix3} for the corresponding expressions. Therefore, the dominant contribution to the fine-tuning remains associated with PBH formation, rather than the specific choice of the particle DM component or the details of the cosmological history.

\section{Alternative PBH formation mechanisms}
The standard PBH formation scenario is the gravitational collapse of rare, large-amplitude primordial density perturbations generated during inflation. Within the Press--Schechter framework, the initial PBH abundance depends exponentially on the ratio of the critical collapse threshold $\delta_c$ to the primordial curvature power spectrum $P_\zeta$, making the predicted PBH abundance extremely sensitive to small variations in the underlying inflationary parameters. As a result, even minute parameter shifts can lead to either an overproduction of PBHs or a negligible relic abundance.

Alternative PBH formation mechanisms have been proposed in which PBHs are generated through different macroscopic processes, such as collapsing domain walls or first-order phase transitions. We here check how model dependent our conclusions to date are against changes in the PBH formation mechanism.

\subsection{PBH formation from first-order phase transition bubble collisions}

A first-order phase transition (FOPT) proceeds through the nucleation of true-vacuum bubbles within the surrounding false vacuum. After nucleation, these bubbles rapidly expand and eventually collide with one another as the phase transition completes. While most collisions simply convert the remaining false vacuum into the true vacuum, rare collision configurations can trap small regions of false vacuum. These trapped regions subsequently collapse under their own gravity, leading to the formation of PBHs, see e.g.~Refs.~\cite{hawking1982bubble, kodama1982abundance, jung2024primordial}.

The PBH abundance is therefore controlled by two ingredients: the probability of producing sufficiently large bubbles and the probability that their collision results in gravitational collapse. In the exact treatment of Ref.~\cite{jung2024primordial}, these effects are computed through integrals over the bubble nucleation history and the collision probability. Refs.~\cite{gouttenoire2024primordial,arteaga2025gravitational} instead provides a convenient semi-analytic approximation to the numerical calculation, showing that the collapse probability is exponentially suppressed,

\begin{equation}
\label{eqPcoll}
\mathcal{P}_{\rm coll}
\simeq
\exp\!\left[
-a
\left(\frac{\beta}{H}\right)^b
(1+\delta_c)^{c\beta/H}
\right]
\;\propto\;
\exp\!\left[
-\frac{\beta}{H}
(1+\delta_c)^{\beta/H}
\right],
\end{equation}
where $a, b, c$ are fitted parameters of order one, and the final proportionality keeps only the dominant exponential dependence for the purpose of estimating the local fine tuning.

Applying \cite{gouttenoire2024primordial} to ultra-light PBHs, the present-day PBH Planckian relics as DM fraction, 
$f_{\rm relic}\equiv\rho_{\rm PBH,0}/\rho_{\rm DM,0}$,
is related to the collapse probability through

\begin{equation}
\label{eq:fPBH}
f_{\rm relic}
=
\frac{\mathcal{P}_{\rm coll}
M_{\rm PBH}
\mathcal{N}_p}
{\Omega_{\rm DM}V_0}
\left( \frac{M_{\rm Pl}}{M_{\rm PBH}}\right)
\simeq
8\times10^{19}
g_*(T_v)^{-1/4}
\left(\frac{M_{\rm PBH}}{\rm g}\right)^{-3/2}
\mathcal{P}_{\rm coll},
\end{equation}
where we used $V_0 = \frac{4 \pi}{3}\rho_c H_0^{-3}$. The PBH mass is approximately

\begin{equation}
\label{eq:MPBH}
M_{\rm PBH}^{\rm bubble} \simeq M_{\rm PBH}
\simeq
1.7\times10^{32}
g_*(T_v)^{-1/2}
\left(\frac{T_v}{\rm GeV}\right)^{-2}.
\end{equation}

Retaining only the dominant parameter dependence, the PBH relic abundance therefore scales as

\begin{equation}
\label{eq:fPBH_scaling}
f_{\rm relic}
\propto
\frac{
\exp\!\left[
-\frac{\beta}{H}
(1+\delta_c)^{\beta/H}
\right]
}{
M_{\rm PBH}^{3/2}
}, 
\end{equation}
which provides a convenient expression for estimating the local fine-tuning measure. Fixing $m_{\chi} = 100~{\rm GeV}$ as a benchmark, we consider the tripartite DM scenario in which the total DM abundance is equally divided between thermal WIMPs, which thermalize the DM produced by PBH evaporation in radiation domination, and Planck-mass PBH relics. The resulting Barbieri--Giudice fine-tuning measures are
\begin{equation*}
\Delta_{\beta/H} \simeq  32 - 69, 
\qquad
\Delta_{M_{\rm PBH}} \simeq 0.72,
\end{equation*}
for an initial PBH mass in the range
$1~{\rm g} \leq M_{\rm PBH} \leq 10^{6}~{\rm g}$.

Although the abundance remains sensitive to the phase-transition parameters, the dominant tuning arises from a single exponential suppression in the collapse probability, rather than the double exponential dependence associated with primordial density perturbation collapse. Therefore, the bubble-collision mechanism may be less fine-tuned, at least compared to popularly considered single-field inflationary models, but a quantitative comparison requires specifying the underlying phase-transition model and fundamental parameter space.

More recent studies have revisited PBH production from supercooled FOPTs using different theoretical approaches, reaching differing conclusions regarding the efficiency of this production mechanism~\cite{franciolini2026curvature,ai2026reviving,Mansour2026sdx}. The calculation therefore remains a subject of ongoing investigation and lies beyond the scope of this work.

\subsection{Domain walls and network collapse}
Another compelling non-inflationary channel for PBH generation arises from the dynamics of cosmological domain walls. When a discrete symmetry is spontaneously broken in the early Universe, smooth topological defects known as domain walls separate distinct vacuum regions. As the Universe expands, the domain wall network enters a scaling regime where the typical wall size and inter-wall separation grow proportionally to cosmic time ($L \propto t$). Under standard conditions, this network would eventually dominate the Universe's energy budget. However, if explicit symmetry-breaking terms—such as those induced by quantum gravity effects—introduce a bias between the vacua, a volume pressure differential forces the domain wall network to collapse and annihilate. 

During this annihilation era, stochastic fluctuations in the network's geometry occasionally cause large, closed domain walls to detach and form closed spherical shells. If a collapsing wall shell is sufficiently symmetric and falls within its own Schwarzschild radius, it undergoes gravitational collapse into a PBH. 

The probability $\mathcal{F}$ of a domain wall collapsing into a black hole at the annihilation time $t_{\rm ann}$ is highly suppressed. From the numerical and analytical results of Gouttenoire et al.~\cite{gouttenoire2025cosmological}, this suppression factor is modeled as:
\begin{equation}
\label{eq:F_ann}
\mathcal{F}\!\left(r_{\rm ann}^{\rm PBH}\right)
\simeq
\exp\!\left[
-
\frac{a}{\ell^{\,b}}
\left(
\frac{1}{\alpha_{\rm ann}}
\right)^{c/\ell^{\,d}}
\right]
\simeq
\exp\!\left[
-\,\alpha_{\rm ann}^{-1}
\right],
\end{equation}
where $\ell \equiv L/t$ represents a scaling parameter of order one, $a, b, c, d$ are fitting parameters of order one, and $\alpha_{\rm ann}$ is an efficiency parameter. Therefore, the total present-day PBH Planckian relics as DM fraction $f_{\rm PBH}$ from this mechanism is given by:
\begin{equation}
\label{eq:fPBH_dw1}
f_{\rm relic}
\simeq
6\times10^{20}\,
g_*(T_{\rm dom})^{-1/4}
\left(
\frac{M_{\rm PBH}}{\rm{g}}
\right)^{-3/2}
\,
\mathcal{F}\!\left(r_{\rm ann}^{\rm PBH}\right)
\,.
\end{equation}
Equivalently, this can be written to highlight the scaling behavior as:
\begin{equation}
\label{eq:fPBH_dw2}
f_{\rm relic}
\propto
\frac{e^{-1/\alpha_{\rm ann}}}{
M_{\rm PBH}^{3/2}}.
\end{equation}
While this framework bypasses the requirement to fine-tune the initial power spectrum,  the PBH abundance still has an  exponential dependence on the efficiency parameter $\alpha_{\rm ann}$.

\section{Conclusion}
In this work, we investigated the fine-tuning properties of a tripartite dark matter (DM) framework consisting of Planck-mass primordial black hole (PBH) relics, DM particles produced through Hawking evaporation, and an additional independent DM component. We considered thermal freeze-out, freeze-in, and QCD axion DM to determine whether the local fine-tuning of the framework depends on the underlying DM candidate. We find that it does not: the qualitative fine-tuning structure is largely independent of the particle physics realization of the third DM component. 

Instead, the dominant source of fine-tuning arises from PBH formation. For PBHs produced from primordial density fluctuations, the DM abundance is significantly more sensitive to the primordial curvature power spectrum, $P_\zeta$, than to particle masses or interaction strengths because the initial PBH abundance depends exponentially on the amplitude of the primordial fluctuations. This conclusion remains unchanged for all DM candidates considered, including the QCD axion, despite modifications to its relic abundance in non-standard cosmological histories. We further showed that alternative PBH formation mechanisms, such as supercooled first-order phase transitions and collapsing domain walls, do not remove the fine-tuning problem; rather, the exponential sensitivity is transferred to the parameters governing these formation processes, such as $\beta/H$ and $\alpha_{\rm ann}$.

We further showed that early PBH domination substantially alters the cosmological evolution of the dark sector. In this regime, entropy injection from PBH evaporation dilutes pre-existing particle abundances and drives the system toward an attractor-like behavior in which the final relic abundance becomes independent of the initial PBH fraction (provided it was initially large enough). 
Consequently, the apparent fine-tuning associated with $\beta_i$ is removed. Nevertheless, the underlying inflationary sector must still be tuned sufficiently to produce a large enough PBH abundance to trigger the early matter-dominated phase, meaning that the fundamental sensitivity is not removed. 



Finally, we emphasize that our analysis is restricted to the local Barbieri--Giudice measure of fine-tuning and therefore does not include the dependence on the fundamental parameters of the underlying inflationary or particle-physics models. A complete assessment would require propagating the sensitivity to those fundamental parameters, which is beyond this work. We also note that our conclusions are largely independent of the specific PBH evaporation mechanism. Alternative scenarios, such as evaporation modified by the memory-burden effect \cite{Dvali2020wqi}, would mainly shift the allowed PBH mass range and abundance required to satisfy cosmological constraints, without altering the central conclusion that the dominant fine-tuning originates from the mechanism responsible for PBH formation.



\begin{center}
    {\bf Acknowledgements}
\end{center}

We would like to thank all those who contributed to this work. In particular, we are grateful to Anish Ghoshal and Philippa Cole for their valuable help and discussions. CB thanks Stefano Profumo for helpful discussions about his related paper which was released while ours was in preparation. Amirah is supported by the Government of KSA through Taibah University. CB is supported by STFC grants ST/X001040/1 and ST/X000796/1.

\printbibliography

\newpage
\pagenumbering{roman}
\medskip
\section{Appendex}

\section*{A1. Dependence on the ratio of $f_{\rm relic}$ and $f_{\chi}$}
\label{Appendix2}
\begin{figure}[ht]
    \centering
    \includegraphics[width=1\linewidth]{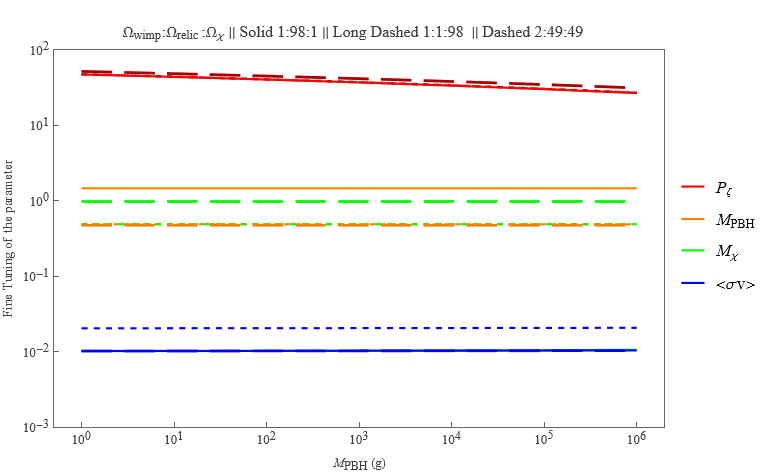}
    \caption[...]{Barbieri--Giudice fine-tuning measures for the primordial power spectrum $P_{\zeta}$ (red), PBH mass $M_{\rm PBH}$ (orange), DM mass $m_{\chi}$ (green), 
    and annihilation cross section $\langle\sigma v\rangle$ (blue) as functions of the PBH mass in grams, for different choices of the tripartite DM composition. 
    While the relative contributions of the three DM components modify the numerical values of the fine-tuning measures, the overall hierarchy remains unchanged: the fine-tuning associated with the primordial power spectrum is consistently the largest over the entire parameter space considered. Thus, our main conclusion is robust against the assumed partitioning of the tripartite DM abundance.}
    
    \label{figRandP}
\end{figure}

In our analysis, we assume equal contributions from PBH relics and PBH-evaporated DM particles, $f_{\rm relic}=f_{\chi}$. To verify that this choice does not affect our conclusions, we repeat the fine-tuning analysis for different relative fractions of the tripartite DM components. The resulting fine-tuning distributions are shown in \cref{figRandP}.

The robustness of the fine-tuning hierarchy allows us to adopt $f_{\rm relic}=f_{\chi}$ as a representative benchmark throughout the main analysis, without loss of generality.

\section*{A2. Non-standard cosmology scenarios for QCD Axion -- formulas}
\label{Appendix3}

In this appendix, we collect the expressions for the axion relic abundance in non-standard cosmological histories following \cite{venegas2021relic}. Although early matter domination and kination modify the expansion history of the Universe and the onset of axion oscillations, the resulting relic abundance retains a simple power-law dependence on the relevant parameters, primarily the axion mass $m_a$ and the characteristic temperature scales describing the cosmological evolution, such as $T_{\rm end}$. This behaviour allows us to assess the impact of different cosmological histories on the fine-tuning without introducing additional exponential sensitivity.

\subsubsection*{Early matter domination ($\mu < 4$)}
\textbf{Region 1:} $R_{\mathrm{osc}} \ll R_{\mathrm{eq}}$
\[
\Omega_{R1} =
\begin{cases} 
\Omega_{a,-3/2}^{sc}
\left(\dfrac{\mu^2}{4}\right)^{-3/\mu}
\left(\dfrac{T_{\mathrm{end}}}{T_{\mathrm{eq}}}\right)^{12/\mu-3},
& \text{for } m_a \lesssim m_{R1},
\\[1.2em]
\Omega_{a,-7/6}^{sc}
\left(\dfrac{\mu^2}{4}\right)^{-3/\mu}
\left(\dfrac{T_{\mathrm{end}}}{T_{\mathrm{eq}}}\right)^{12/\mu-3},
& \text{for } m_a \gtrsim m_{R1}.
\end{cases}
\]
\vspace{1em}
\textbf{Region 2:} $R_{\mathrm{eq}} \ll R_{\mathrm{osc}} \ll R_c$
\[
\Omega_{R2} \simeq
\begin{cases}
\Omega_{a,-3/2}^{sc}
\left(\dfrac{T_{\mathrm{end}}^2}{m_a M_{\rm Pl}}\right)^{\frac{3}{2\mu}(4-\mu)},
& \text{for } m_a \lesssim m_{R2},
\\[1.2em]
\Omega_{a,-7/6}^{sc}
\left(\dfrac{\mu^2}{4}\right)^{-3/\mu}
\left(\dfrac{T_{\mathrm{end}}}{T_{\mathrm{eq}}}\right)^{\frac{3}{\mu}(4-\mu)}
\left[
\dfrac{T_{\mathrm{eq}}^7}
{\left(T_{\mathrm{QCD}}^4\,m_a M_{\rm Pl}\right)^{7/6}}
\right]^{\frac{4-\mu}{\mu+8}},
& \text{for } m_a \gtrsim m_{R2}.
\end{cases}
\]
\vspace{1em}
\textbf{Region 3:} $R_c \ll R_{\mathrm{osc}} \ll R_{\mathrm{end}}$
\[
\Omega_{R3} =
\begin{cases}
\Omega_{a,-3/2}^{sc}
\left(\dfrac{2}{\mu}\right)^{6/\mu}
\left(\dfrac{T_{\mathrm{end}}^2}{m_a M_{\rm Pl}}\right)^{\frac{3}{2\mu}(4-\mu)},
& \text{for } m_a \lesssim m_{R3},
\\[1.2em]
\Omega_{a,-7/6}^{sc}
(8-\mu)^{\frac{\mu+6}{24}}
\left(
\dfrac{T_{\mathrm{end}}^6}
{T_{\mathrm{QCD}}^4\,m_a M_{\rm Pl}}
\right)^{3/\mu-2/3},
& \text{for } m_a \gtrsim m_{R3}.
\end{cases}
\]
\subsubsection*{Kination ($\mu > 4$)}
\[
\Omega_a =
\begin{cases}
\Omega_{a,-3/2}^{sc}
\left(
\dfrac{T_{\mathrm{end}}^2}{m_a M_{\rm Pl}}
\right)^{\frac{12-3\mu}{2\mu}},
& m_a \lesssim m_{\mathrm{QCD}},
\\[1.2em]
\Omega_{a,-7/6}^{sc}
\left[
\dfrac{
\left(T_{\mathrm{QCD}}^4\,m_a M_{\rm Pl}\right)^{1/6}
}
{T_{\mathrm{end}}}
\right]^{7\frac{\mu-4}{\mu+8}},
& m_a \gtrsim m_{\mathrm{QCD}}.
\end{cases}
\]

The above expressions determine the dependence of the axion abundance on the underlying cosmological parameters used in the fine-tuning analysis - see \cref{tabNSCaxion}. Despite the modified expansion histories, the axion relic density remains a smooth power-law function of these parameters. Consequently, non-standard cosmological scenarios can significantly modify the axion abundance while not introducing a new source of strong sensitivity comparable to the exponential dependence of the PBH formation abundance on the primordial power spectrum amplitude $P_{\zeta}$.
\begin{table}[ht]
\centering
\renewcommand{\arraystretch}{1.5}
\begin{tabular}{|c|c|c|c|c|}
\hline
\textbf{Cosmic History} & \textbf{Region} & \textbf{$p$} & \textbf{Fine-Tuning} & \textbf{Order} \\ \hline
{\textbf{eMD} ($\mu=3$)} & 

{\begin{tabular}[c]{@{}l@{}} \\ 
$\Omega_{R1} \propto T_{\rm{end}}
\begin{cases} 
m_a^{-7/6} & \text{for } m_a \gtrsim m_{R1} \\
m_a^{-3/2} & \text{for } m_a \lesssim m_{R1}
\end{cases}$
\end{tabular}} 

& $T_{\text{end}}$ & $1.0$ & $\mathcal{O}(1)$ \\ \cline{3-5} 
 & & $m_a$ & $1.2$ and $1.5$ & $\mathcal{O}(1)$ \\ \cline{2-5} 
 & 

{\begin{tabular}[c]{@{}l@{}} \\ 
$\Omega_{R2} \propto T_{\rm{end}}
\begin{cases} 
m_a^{-14/11}  & \text{for } m_a \gtrsim m_{R2} \\
m_a^{-2} & \text{for } m_a \lesssim m_{R2}
\end{cases}$
\end{tabular}} 

 & $T_{\text{end}}$ & $1.0$ & $\mathcal{O}(1)$ \\ \cline{3-5} 
 & & $m_a$ & $1.3$ and $2.0$ & $\mathcal{O}(1)$ \\ \cline{2-5} 
 & 

{\begin{tabular}[c]{@{}l@{}} \\ 
$\Omega_{R3} \propto 
\begin{cases} 
T_{\rm{end}}^2 \,m_a^{-3/2} & \text{for } m_a \gtrsim m_{R3} \\
T_{\rm{end}} \, m_a^{-2} & \text{for } m_a \lesssim m_{R3}
\end{cases}$
\end{tabular}} 

& $T_{\text{end}}$ & $2.0$ and $1.0$ & $\mathcal{O}(1)$ \\ \cline{3-5} 
 & & $m_a$ & $1.5$ and $2.0$ & $\mathcal{O}(1)$ \\ \cline{2-5} 
\hline
{\textbf{Kination} ($\mu=6$)} & 

{\begin{tabular}[c]{@{}l@{}} \\ 
$\Omega_{QCD} \propto 
\begin{cases} 
T_{\rm{end}}^{-1} \,m_a^{-1} & \text{for } m_a \gtrsim m_{QCD} \\
T_{\rm{end}} \, m_a^{-2} & \text{for } m_a \lesssim m_{QCD}
\end{cases}$
\end{tabular}} 

& $T_{\text{end}}$ & $1.0$ & $\mathcal{O}(1)$ \\ \cline{3-5} 
 & & $m_a$ & $1.0$ and $2.0$ & $\mathcal{O}(1)$ \\ \cline{2-5} 
\hline
\end{tabular}
\caption{Barbieri--Giudice fine-tuning sensitivities of the QCD axion relic abundance in non-standard cosmological scenarios. The values are obtained from the analytical expressions of Ref.~\cite{venegas2021relic}. Here $\mu=3(1+\omega)$ parametrizes the equation of state of the dominant component, with $\mu=3$ for early matter domination and $\mu=6$ for kination. 
}
\label{tabNSCaxion}
\end{table}


\section*{A3. Dark radiation and \texorpdfstring{$\Delta N_{\rm eff}$}{Delta Neff} from PBH evaporation}
\label{Appendix4}

In this appendix we summarize the calculation of the contribution to the effective number of relativistic species, $\Delta N_{\rm eff}$, arising from the Hawking emission of dark radiation from PBHs \cite{hooper2019dark}. If PBHs dominate the energy density of the Universe prior to evaporation, the energy density transferred into dark radiation is given by
\begin{equation*}
\rho_{\rm DR}
=
\rho_{\rm BH}
\frac{g_{{\rm DR},H}}
{g_{*,H}},
\end{equation*}
where $g_{{\rm DR},H}$ denotes the Hawking-weighted number of dark-radiation degrees of freedom and
\begin{equation*}
g_{*,H}
=
\sum_i g_{i,H}
\end{equation*}
is the total Hawking-weighted number of accessible particle degrees of freedom.

Following \cite{hooper2019dark}, the resulting contribution to the effective number of neutrino species is
\begin{equation}
\Delta N_{\rm eff}
=
\left(
\frac{g_{{\rm DR},H}}
{g_{*,H}}
\right)
\left(
\frac{g_{*S}(T_{\rm EQ})}
{g_{*S}(T_{\rm RH})}
\right)^{1/3}
\left(
\frac{g_{*S}(T_{\rm EQ})}
{g_{*}(T_{\rm EQ})}
\right)
\left[
N_\nu
+
\frac{8}{7}
\left(\frac{11}{4}\right)^{4/3}
\right],
\label{eq:DeltaNeff_dom}
\end{equation}
where $T_{\rm RH}$ denotes the reheating temperature after PBH evaporation and $T_{\rm EQ}$ corresponds to matter-radiation equality.

Using $N_\nu = 3.046$, and $g_{*}(T_{\rm EQ}) \simeq g_{*S}(T_{\rm EQ}) \simeq 3.38$, \cref{eq:DeltaNeff_dom} simplifies to
\begin{equation}
\Delta N_{\rm eff}
\simeq
8.68
\left(
\frac{g_{{\rm DR},H}}
{g_{*,H}}
\right)
\left(
\frac{g_{*S}(T_{\rm EQ})}
{g_{*S}(T_{\rm RH})}
\right)^{1/3} \simeq 13.71
\frac{g_{{\rm DR},H}}
{g_{*,H} \,\,\, g_{*S}(T_{\rm RH})^{1/3}}.
\end{equation}

For temperatures above the electroweak scale one may take $g_{*S}(T_{\rm RH})
\simeq 106.75$. For a single QCD axion, treated as a real scalar degree of freedom, $g_{{\rm DR},H} = g_{{\rm a},H} = 1.82$ (for spin $=0$). For sufficiently hot PBHs, where all Standard Model degrees of freedom are emitted, $g_{*,H}\simeq108$, one finds
\begin{equation}
\Delta N_{\rm eff}
\simeq
0.049
\times
\begin{cases}
1,
&
\beta_i
\ge
\beta_c
\\[0.4cm]
\displaystyle
\beta_i \left(\frac{t_{\rm ev}}{t_i}\right)^{1/2},
&
\beta_i
<
\beta_c.
\end{cases}
\end{equation}

\section*{A4. Temperature of the Universe at PBH formation and evaporation}
\label{Appendix1}
In this appendix, we derive the background radiation temperature at the time of PBH formation, $t_i$, and at the end of PBH evaporation, $t_{\rm ev}$, assuming a radiation-dominated Universe. During radiation domination, the temperature scales with time as $T\propto t^{-1/2}$. Therefore, the background temperature at PBH formation can be obtained by evolving the Universe temperature from the epoch of BBN backwards to $t_i$:
\begin{equation}
\label{eqTi}
    T_i =
    \left(\frac{t_{\rm BBN}}{t_i}\right)^{1/2} T_{\rm BBN}
    =
    1.04\times10^{29}
    \left(\frac{M_{\rm PBH}}{\rm g}\right)^{-1/2}
    \,{\rm K},
\end{equation}
where we have used $t_{\rm BBN}\simeq1~{\rm s}$ and $T_{\rm BBN}\simeq1~{\rm MeV}\simeq1.16\times10^{10}~{\rm K}$. The dependence on the PBH mass follows from the relation between the PBH formation time and its initial mass during radiation domination.

The lifetime of a PBH with initial mass $M_{\rm PBH,i}$, evaporating into $g$ relativistic degrees of freedom, is given by
\begin{equation}
    t_{\rm ev}\simeq
    \frac{4\times10^{-26}}{g}
    \left(\frac{M_{\rm PBH,i}}{\rm g}\right)^3
    {\rm s},
\end{equation}
where the numerical coefficient includes the standard graybody-factor correction.

To determine the background temperature at the time of evaporation, we use the radiation-dominated Friedmann equation,
\begin{equation}
    t(T)=\frac{1}{2H}
    =
    \frac{1}{2}
    \left(
    \frac{8\pi G\rho_{\rm rad}}{3}
    \right)^{-1/2},
\end{equation}
where the radiation energy density is
\begin{equation}
    \rho_{\rm rad}
    =
    \frac{\pi^2}{30}g_*
    T^4
    \left(\frac{k_B^4}{\hbar^3c^5}\right).
\end{equation}

Substituting the PBH lifetime into the radiation-era temperature--time relation gives the background temperature when evaporation completes:
\begin{equation}
\label{eqTevap}
    T_{\rm evap}
    \simeq
    2.8\times10^{23}
    \left(\frac{g_*}{106.75}\right)^{-1/4}
    \left(\frac{g}{106.75}\right)^{1/2}
    \left(\frac{M_{\rm PBH}}{\rm g}\right)^{-3/2}
    \,{\rm K}.
\end{equation}
Here, we have neglected any increase in the radiation temperature due to the energy injected by PBH evaporation itself, and $T_{\rm evap}$ should therefore be interpreted as the background radiation temperature at the evaporation time. 

\clearpage
\phantomsection

\end{document}